# Automatic Deep Learning System for COVID-19 Infection Quantification in chest CT

College of Engineering and Technology, American University of the Middle East, Kuwait
Corresponding author: Omar Alirr (omar.alirr@aum.edu.kw).

*Abstract*—Coronavirus Disease spread globally and infected millions of people quickly, causing high pressure on the health-system facilities. PCR screening is the adopted diagnostic testing method for COVID-19 detection. However, PCR is criticized due its low sensitivity ratios, also, it is time-consuming and manual complicated process. CT imaging proved its ability to detect the disease even for asymptotic patients, which make it a supportive screening tool to PCR. In addition, the appearance of COVID-19 infections in CT slices, offers high potential to support in disease evolution monitoring using automated infection segmentation methods. However, COVID-19 infection areas include high variations in term of size, shape, contrast and intensity homogeneity, which impose a big challenge on segmentation process. To address these challenges, this paper proposed an automatic deep learning system for COVID-19 infection areas segmentation. The system start prepare the region of interest by segmenting the lung organ, which then undergo edge enhancing diffusion filtering (EED) to improve the infection areas contrast and intensity homogeneity. The proposed FCN is implemented using U-net architecture with modified residual block with concatenation skip connection. To demonstrate the generalization and effectiveness of the proposed system, it is trained and tested using many 2D CT slices extracted from diverse datasets from different sources. The proposed system is evaluated using different measures and achieved dice overlapping score of 0.961 and 0.780 for lung and infection areas segmentation, respectively.

*Keywords— COVID-19 infection, deep learning, segmentation, chest CT*

## I. INTRODUCTION

The pandemic of novel coronavirus disease (COVID-19) is affecting 213 countries and territories around the world, and with 2 international conveyances [1]. According to the statistics coming from the COVID-19 dashboard of Centre for Systems Science and Engineering (CSSE) at Johns Hopkins University, more than 19,172,505 identified cases of COVID-19 have been reported (the number is increasing), including 716,327 deaths [2]. The high contagiousness of COVID-19 is the reason of rapid increase of the confirmed cases. COVID-19 leads to extreme respiratory problems with different symptoms include fever, cough and fatigue. These symptoms can develop into severe pneumonia especially to people with weakened immune systems [3].

The Reverse Transcription Polymerase Chain Reaction (RT-PCR) is the most commonly used test to detect the viral RNA using nasopharyngeal swab. However, RT-PCR test is reported by different studies to have high false negative rates, where a repeated test is needed for accurate diagnosis. In addition, RT-PCR test has availability limitations due to the shortage of manufacturing material, also the testing process is time consuming that limits the rapid and accurate screening [4][5].

The computed tomography (CT) is an alternative solution to RT-PCR for COVID-19 screening, where high proportions of CT scans were obtained from the infected patients. Compared to other types of tests, CT scanning is considered as a promising and efficient alternative tool for the detection and control of COVID-19 disease. CT imaging has been recommended for COVID-19 diagnosis, specifically, the chest CT screening has been used as a routine diagnostic tool for pneumonia [4]. Chest CT scanning has demonstrated effectiveness in coronavirus disease diagnosis, including follow-up assessment and disease progression monitoring [6][7].

Diagnostic studies using CT screening on COVID-19 disease patients state that the infection areas may appear in the CT scans before the appearance of the disease symptoms. Therefore, for asymptomatic patients, these COVID-19 infection areas can be detected by observing the ground glass opacity (GGO) and pulmonary consolidations signs, which could appear at different stages of the disease [8][9].

Visual CT imaging analysis can help in COVID-19 disease diagnosis by proposing approaches to identify the predominant patterns of the infections like ground glass opacity (GGO) and pulmonary consolidations. Systems in this field support three different image processing tasks. First, CT classification, where the patient is classified to have the disease or not[10][11]. Second, the disease infection detection, where the infection areas are highlighted by bounding boxes. Third task is the infection area segmentation and disease burden calculation by applying classification at pixel level [12].

Manual segmentation of COVID-19 infections is tedious and time-consuming process. In addition, it greatly depends on the skills of the physician or doctor who perform the segmentation task [13][14]. For COVID-19 infection segmentation, the automated approach is desirable, as it is, ideally, more objective and removes dependence on human skills. Recently, due to the advancement in computer vision, the development of deep fully convolutional networks (FCN) enhanced the performance of the semantic segmentation, which leads to outperform other competitors in the field of medical imaging [15]–[17].

General FCN focuses its task on image classification, where input is an image and output is one label. However, in COVID-19 chest CT analysis, it requires, beside the classification, to localize and segment the area of abnormality[16][18]. Researchers started to use FCN for COVID-19 disease to help clinicians and radiologists in diagnosis and prognosis tasks, which succeeded to improve accuracy and reduce the time of inspection [19][20].

Deep learning systems have been proposed by many researchers to help in combating the high spread of COVID-19 disease. Most of the recent proposed deep learning systems focus on detecting (classifying) the patients infected by COVID-19 disease using CT screening, that is due to the availability of CT scans, which are many, and it does not require radiologist annotations, which are very rare [7][21].

As an example for classification COVID-19 diagnosis systems, COVID-Net which is introduced by authors in [22], it is a deep neural network tailored for the detection of COVID-19 cases from chest X-ray images that are open source and available to the general public. Other study proposed 3D deep learning system trained using pulmonary CT images to distinguish COVID-19 pneumonia from Influenza-A viral pneumonia and healthy cases [23]. Another weakly-supervised deep learning-based software system was developed using 3D CT volumes to detect COVID-19 [24].

Many deep learning systems have been proposed to assist COVID-19 diagnosis in clinical practice, however few of them are related to infection delineation from CT scans [25]–[29]. Most of these proposed techniques use U-net FCN implementation as backbone in their approaches. some other works proposed their own COVID-19 oriented deep networks [27][30]. The common challenge for most of the proposed methods is the insufficient labeled CT scans for deep networks training, which cannot be available in short time. As the process of annotating the infection areas is time consuming, expensive and depends on the radiologist expertise[12].

## II. METHOD

### A. Overview of the Proposed framework

Figure 1 presents the flowchart of the proposed segmentation system; it consists of two main steps that are applied sequentially. The first step is the lung segmentation step from the plain chest CT slices, and then it is followed by the COVID-19 infection. The proposed framework depends on using two cascaded fully convolutional networks (FCNs). The first FCN is built to segment the lung organ which is used as region of interest (ROI) to focus and segment the COVID-19 infection areas using the second FCN. The two constructed FCNs are analyzed and trained using diverse datasets form different public sources.

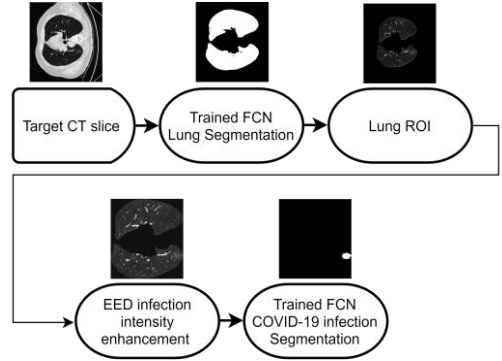

FIGURE 1. Flowchart of COVID-19 infection segmentation pipeline

### B. COVID-19 Datasets

In this paper, publically available COVID-19 chest CT datasets, which are collected from different sources, are used to train and test the proposed FCN networks. All CT datasets were labeled with required classes for training: background, lung organ and COVID-19 infection[31].

The first open access COVID-19 dataset is collected by the Italian Society of Medical and Interventional Radiology (SIRM). The COVID-19 CT segmentation dataset consists of 100 axial slices that contains lung infections segmented by radiologist, where the segmented slices are extracted from more than 40 diverse COVID-19 infected CT scans [32]. The second group of public datasets is from Radiopedia, it consists of 9 COVID-19 chest volumetric CT with corresponding ground-truth. About 373 slices of the whole 9 datasets have been diagnosed as positive and delineated by a radiologist [13].

A third newly released public dataset is collected and presented by Ma et al., which consists of 20 annotated COVID-19 chest CT, 10 CT scans from Coronacases Initiative and another 10 CT scans from Radiopedia [13]. These CT scans are freely accessible with CC BY-NC-SA license, and all the 20 COVID-19 CT scans are labeled by two radiologists and verified by an experienced radiologist.

### C. COVID-19 Infections appearance Enhancement

Different challenges confront the segmentation of infection areas inside the lung ROI, like the blurry edges and intensity inhomogeneity inside infection areas. The COVID-19 infection area has low contrast in the chest CT images; it does not have a clear boundary from the surrounding tissues. In addition, infection areas have high variability in term of texture, size and position in CT slices To improve the detection of the infection areas, the segmented lung (ROI) from the CT image is enhanced using tensor-based Edge Enhancing Diffusion (EED) filtering [33]. EED filtering uses diffusion tensor to adapt the diffusion based on the image structure. EED filter helps to enhance the contrast, filter the noise to improve intensity homogeneity, and preserves the boundaries of the shape [33].

To improve the detection and segmentation of the COVID-19 infection areas, EED filtering is used to increase the contrast of infection areas by enhancing the intensity homogeneity inside these areas and preserves the boundaries with respect to the lung parenchyma. This step aims to improve the FCN training process to extract and learn the main features that differentiate the infection areas from the surrounding tissues. Figure 2 shows the effect of the EED step to the raw medical CT slice.

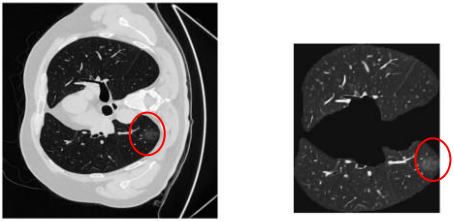

**FIGURE 2.** EED enhancement of COVID-19 infection areas.

*D. Network Architecture: ResDense FCN*

In this work, two cascaded deep FCNs are connected sequentially to segment the lung organ and then the COVID-19 infection areas. The backbone of the proposed FCN network is an adjustment of U-net architecture [34], with 5 levels as shown by figure 4. The U-net consists of an encoding path and decoding path. At each level of the encoding, three operations are applied: convolution, activation function (ReLU) and batch normalization. These operations are applied two times consecutively in each level block, which is followed by max-pooling operation before moving to the next level. The kernel size is 3x3 for convolutions and 2x2 for the max-pooling. The resolution of the feature is reduced to the half after each level.

The decoding path of the network recovers the original input size by applying same sequence of operation (conv, ReLU, BN) but replacing the max-pooling with up-sampling at each level. In addition, the corresponding feature from the encoding path is concatenated to the input of each decoding level. The last level in the decoding path ends with 1x1 convolution with sigmoid activation function to classify the feature map using dice coefficient metric and generate the final binary prediction map.

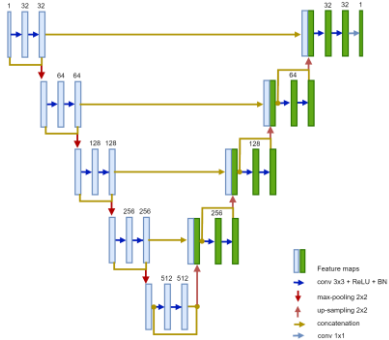

**FIGURE 4.** The proposed FCN

The increase of network depth in FCN is inevitable, however it leads to the vanishing gradient problem, as more layers are stacked together vanish and wash out the gradient information, which slows down the training and degrades the performance. Different deep network architectures were proposed to deal with this problem, however, the DensNet and ResNet are considered as breakthroughs in term of performance.

In DensNet, each layer is connected to all forward layers, where the feature maps generated from different filter sizes are concatenated from previous layers, which make the model much thicker as channels are joined after every convolution operation [35]. On the other hand, that doesn't happen in ResNet as the addition operation is used to merge the previous input identity with output feature map. In ResNet block, a shortcut (skip connection) from the input of the block (identity) bypass the stacked layers and attach with output feature of the block [36], figure 5 explains the difference in connections for (a) Residual block, and (b) Dense block

DensNet aims to ensure maximum information to flow between layers in the network by combining the features through concatenating them instead of summation as on ResNet. Therefore, the DensNet is considered as memory hungry networks, as the back-propagation requires storing the entire layers outputs, which costs more memory and runs slow. On the other hand, the addition of tensors is the idea of ResNet, however it has been argued that the direct addition of feature maps harms the gradient flow through the network, as it sums up the features values.

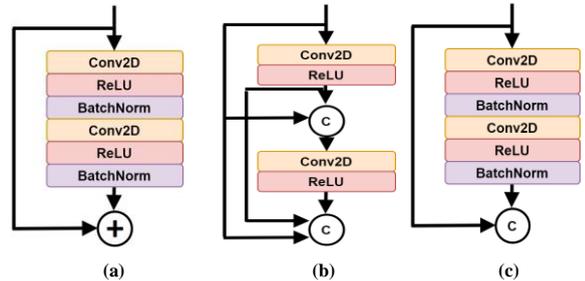

**FIGURE 5.** (a) Residual block (b) Dense block (c) ResDense Block

Therefore, the concatenation operation is preferred as it preservers the feature maps, while the summation corrupts the feature maps for both the convolution operation and the source of the skip connections. The main contribution in the proposed network architecture is the ResDense block as shown in figure 5 (c). In the proposed ResDense, the dense connections (concatenation) are used between residual blocks rather than convolution layers. In terms of feature maps flow and memory, the proposed ResDense block refines the feature values sufficiently by having residual blocks and memorizes the refined feature values intermittently by dense connections between residual blocks.

In this work, the proposed FCN network's encoder and decoder paths depend on ResDense blocks. Each level in the contracting and expanding paths of the proposed network is built using ResDense block as shown in figure 4. Hence, the depth size of the feature map is doubled and concatenated with block input at the end of each level in the encoding path.

*E. Network implementation and training*

The proposed network is trained with resized annotated 2D slices with patch size of (256 × 256). All CT slices were normalized patch-wise using zero mean and unit variance normalization. The ResDense FCN architecture is implemented using Keras[1] with the TensorFlow backend.

Two ResDense networks were built to segment both the lung and then the infection areas inside it, sequentially. The network is trained and the parameters are updated using Adam optimizer with learning rate set to 0.0001. Both networks are trained with 20 epochs, with batch size 32, 16 for lung and infections segmentations, respectively. The soft dice coefficient loss was used to update the model parameters and to monitor the network training convergence. Batch normalization (BN) layer is used in the network design, which helps to avoid unrealistic increase or decrease of the generated values among network layers. The final layer for the network utilizes pixel-wise sigmoid activation function.

Due to imbalance class distribution of the lung tissue and the COVID-19 infections, many steps have been taken to improve the performance of the trained network. First, the soft dice loss metric is used in the training process to measure the overlap between the ground-truth patches and the area labeled as infection by the network inside the lung ROI. In addition, the second FCN network is trained only inside the lung ROI, to learn features that discriminate COVID-19 infections from lung tissues background only. Therefore, the training patches are extracted by cropping the lung part only from the CT slice. Besides that, the used training patches are ensured to have the corresponding mask, and to exclude patches without annotation mask from the training process.

*F. Performance Measures*

To quantitatively evaluate the proposed system, a group of performance measures are used to assess the segmentation of COVID-19 infection and lung organ from CT scans. First, the Dice coefficient (DSC), it is an overlap measure that computes the ratio between the correctly segmented class with respect to the average size between the segmentation output and the ground truth. Dice metric (DSC) is given by equation 1, where TP (true positive) and TN (true negative) represent the number of voxels correctly classified in the segmented class and in background, respectively. FN (false negative) and FP (false positive) represent the number of voxels that nor correctly classified as segmented class neither as background.

$$DSC = \frac{2*TP}{2*TP + FN + FP} \quad (1)$$

The second measure is sensitivity, which finds the ratio of the correctly segmented class voxels (TP) compared to the ground truth (TP+FN), as given by Equation 2. The third measure is the specificity, which measure the ratio of the correctly segmented non-class voxels (TN) over the total number of non-class voxels, the specificity is shown by Equation 3.

$$\text{Sensitivity} = \frac{TP}{TP + FN} \quad (2)$$

$$\text{Specificity} = \frac{TN}{TN + FP} \quad (3)$$

### III. RESULTS AND DISCUSSION

Datasets are divided into two groups, training and testing groups. The training group consists of CT scans from Radiopedia and Coronacases Initiative datasets, while the testing dataset contains the SRIM dataset. The proposed FCN networks are trained with 2D labeled slices that are extracted from the training dataset group, where 3,686 and 2,216 slices are used for lung organ and COVID-19 infections segmentations, respectively.

The proposed method is evaluated qualitatively and quantitatively on the diverse test datasets (SIRM dataset). The trained networks showed an impressive performance for both lung and COVID-19 infection segmentation. Figure 6 shows the visual segmentation comparison of the segmented results (red) for three different examples with the corresponding ground-truths (green). While, figure 7 shows the visual comparison for COVID-19 infection segmentation (red) and corresponding ground-truths (green). From the qualitative evaluation presented by figure 7, it is observed that the segmented infections areas (red) revealed the strong performance of the proposed system.

Quantitative evaluation proved the high performance of the proposed system as explained by the achieved values of the performance measures in table 1. The achieved DSC values are 0.96 for lung segmentation and 0.78 for COVID-19 infections. In addition, the system achieved sensitivity and specificity of 0.93 and 0.99 for lung segmentation, and 0.82 and 0.95 for infection segmentation, respectively.

TABLE I
MEASURES VALUES FOR LUNG AND COVID-19 INFECTION SEGMENTATION

| Task | DSC | Sensitivity | Specificity |
|---|---|---|---|
| Lung | 0.961±0.018 | 0.932±0.007 | 0.996±0.028 |
| Infection | 0.780±0.144 | 0.822±0.155 | 0.951±0.049 |

---

[1] https://keras.io/

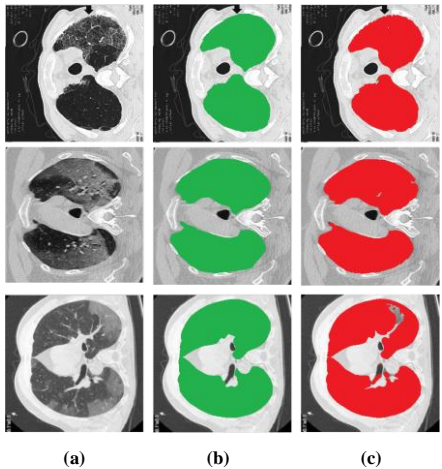

**FIGURE 6. Visual Lung segmentation comparison: (a) axial slice (b) Ground-truth (c) segmented lung organ**

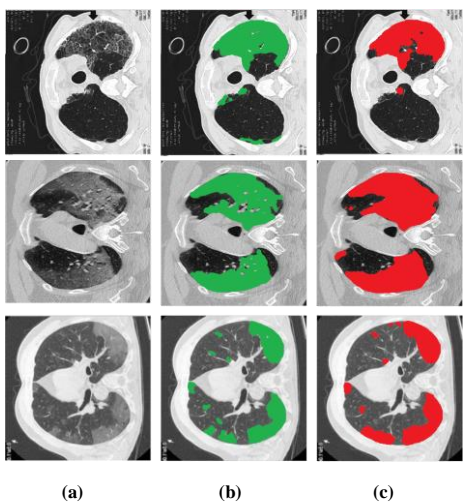

**FIGURE 7. Visual COVID-19 infections segmentation comparison: (a) axial slice (b) Ground-truth (c) segmented COVID-19 infection areas**

For further evaluation, a comparative study is carried to assess the proposed system with respect to other COVID-19 segmentation approaches. However, it is difficult to establish a fair comparison due lack of common public datasets. Table 2 presents the proposed method along with other state of art COVID-19 infection segmentation. Some of these methods used publically available datasets, while other works used their own private datasets for their system analysis. From table 2, the proposed method succeeded to achieve high DSC value compared to other listed methods.

The lack of public annotated training datasets is the main challenge for COVID-19 deep learning segmentation systems, especially with the hungry deep FCNs that use 3D patches for training [28][30]. Works that are based on limited datasets could suffer from over-fitting, as number of training datasets is few that lead to biased systems lack the ability of generalization.

**TABLE II**
**COMPARISON WITH OTHER PROPOSED METHOD FOR COVID-19 INFECTION SEGMENTATION**

| Method | DSC | Sensitivity | Specificity |
|---|---|---|---|
| Yan et al. [30] | 0.726 | 0.751 | NA |
| Fan et al. [27] | 0739 | 0.725 | 0.960 |
| Qiu et al. [37] | 0.773 | 0.830 | 0.974 |
| Ma et al. [13] | 0.673 | NA | NA |
| **The proposed method** | **0.780±0.41** | **0.822±0.15** | **0.951±0.05** |

However, some authors tried to deal with limited datasets by using cross-validation approach to generate multiple folds from the same small dataset, where each generated fold is used for both training and testing [13][28]. Ma et al. proposed U-net based deep learning system, which is trained using 20 CT scans, 10 cases from the Coronacases Initiative and 10 cases from Radiopedia. They used 20% of the dataset (4 cases) for training and 80% of the dataset (the remaining 16 cases) are used for testing [13]. Similarly, Muller et al. used the same dataset used by Ma et al., but with inverted percentages to train their 3D deep networks ( 80% for training and 20% for testing) [28].

Qiu et al. used SRIM dataset which consists of 100 axial CT images from 60 patients with COVID-19 to train their specific COVID-19 MiniSeg system. Authors supposed that even if the dataset is small but it is diverse with each patient contributing 1.6 axial CT images. They randomly chose 60 images for training models, and another 40 CT images for performance evaluation[37]. On the other hand, some works succeeded to use plenty of their own datasets (private) to train deep learning systems. Yan et al. trained and tested their specifically designed 3D deep system, COVID-SegNet, using chest CT images from 861 patients with confirmed COVID-19, which are annotated by experts [30].

## IV. CONCLUSION

This paper presented a deep learning system for COVID-19 lung infection segmentation in chest CT scans. The constructed FCN utilized U-net architecture as backbone with each level in the encoding and decoding paths is built using proposed ResDense blocks. The feature maps of the infection areas and lung background flow through the network without significant change of their values due to concatenation skip connection in the each ResDense blocks, which improved network learning and enhanced the segmentation performance. Moreover, the system contains EED step to improve the infection areas appearance in CT slices by enhancing their contrast and intensity homogeneity. The qualitative and quantitative evaluation results demonstrate the effectiveness and the ability of the system to segment COVID-19 infection areas from CT images. The system is trained and validated using diverse datasets from different sources, which proved its ability for generalization and be promising tool for automatic analysis of COVID-19 infection detection and in clinical routine.

REFERENCES

[1] "https://www.worldometers.info/coronavirus/.".
[2] Https://coronavirus.jhu.edu/ and accessed: 2020-04-02. Map.html, "'Coronavirus COVID-19 global cases by the center for systems science and engineering at johns hopkins university,'" *"Coronavirus COVID-19 global cases by the center for systems science and engineering at johns hopkins university,."* .
[3] C. Sohrabi *et al.*, "World Health Organization declares global emergency: A review of the 2019 novel coronavirus (COVID-19)," *International Journal of Surgery*, vol. 76. Elsevier Ltd, pp. 71–76, 01-Apr-2020.
[4] T. Ai *et al.*, "Correlation of Chest CT and RT-PCR Testing in Coronavirus Disease 2019 (COVID-19) in China: A Report of 1014 Cases," *Radiology*, p. 200642, Feb. 2020.
[5] Y. Fang *et al.*, "Sensitivity of Chest CT for COVID-19: Comparison to RT-PCR," *Radiology*, vol. 296, no. 2, pp. E115–E117, Aug. 2020.
[6] G. D. Rubin *et al.*, "The Role of Chest Imaging in Patient Management During the COVID-19 Pandemic: A Multinational Consensus Statement From the Fleischner Society," *Chest*, vol. 158, no. 1, pp. 106–116, Jul. 2020.
[7] F. Shi *et al.*, "Review of Artificial Intelligence Techniques in Imaging Data Acquisition, Segmentation and Diagnosis for COVID-19," *IEEE Rev. Biomed. Eng.*, 2020.
[8] A. J. Rodriguez-Morales *et al.*, "Clinical, laboratory and imaging features of COVID-19: A systematic review and meta-analysis," *Travel Medicine and Infectious Disease*, vol. 34. Elsevier USA, p. 101623, 01-Mar-2020.
[9] M.-Y. Ng *et al.*, "Imaging Profile of the COVID-19 Infection: Radiologic Findings and Literature Review," *Radiol. Cardiothorac. Imaging*, vol. 2, no. 1, p. e200034, Feb. 2020.
[10] L. Li *et al.*, "Using Artificial Intelligence to Detect COVID-19 and Community-acquired Pneumonia Based on Pulmonary CT: Evaluation of the Diagnostic Accuracy," *Radiology*, vol. 296, no. 2, pp. E65–E71, Aug. 2020.
[11] C. Butt, J. Gill, D. Chun, and B. A. Babu, "Deep learning system to screen coronavirus disease 2019 pneumonia," *Appl. Intell.*, p. 1, 2020.
[12] X. Ding, J. Xu, J. Zhou, and Q. Long, "Chest CT findings of COVID-19 pneumonia by duration of symptoms," *Eur. J. Radiol.*, vol. 127, p. 109009, Jun. 2020.
[13] J. Ma *et al.*, "Towards Efficient COVID-19 CT Annotation: A Benchmark for Lung and Infection Segmentation," Apr. 2020.
[14] H. Meng *et al.*, "CT imaging and clinical course of asymptomatic cases with COVID-19 pneumonia at admission in Wuhan, China," *J. Infect.*, vol. 81, no. 1, pp. e33–e39, Jul. 2020.
[15] J. Long, E. Shelhamer, and T. Darrell, "Fully convolutional networks for semantic segmentation," in *Proceedings of the IEEE Computer Society Conference on Computer Vision and Pattern Recognition*, 2015, vol. 07-12-June-2015, pp. 3431–3440.
[16] M. H. Hesamian, W. Jia, X. He, and P. Kennedy, "Deep Learning Techniques for Medical Image Segmentation: Achievements and Challenges," *J. Digit. Imaging*, vol. 32, no. 4, pp. 582–596, Aug. 2019.
[17] G. Litjens *et al.*, "A survey on deep learning in medical image analysis," *Medical Image Analysis*, vol. 42. 2017.
[18] W. Li, F. Jia, and Q. Hu, "Automatic Segmentation of Liver Tumor in CT Images with Deep Convolutional Neural Networks," *J. Comput. Commun.*, vol. 03, no. 11, pp. 146–151, Nov. 2015.
[19] G. Litjens *et al.*, "A survey on deep learning in medical image analysis," *Medical Image Analysis*, vol. 42. Elsevier B.V., pp. 60–88, 01-Dec-2017.
[20] J. Bullock, A. Luccioni, K. H. Pham, C. S. N. Lam, and M. Luengo-Oroz, "Mapping the Landscape of Artificial Intelligence Applications against COVID-19," Mar. 2020.
[21] J. Bullock, A. Luccioni, K. H. Pham, C. S. N. Lam, and M. Luengo-Oroz, "Mapping the Landscape of Artificial Intelligence Applications against COVID-19," Mar. 2020.
[22] L. Wang and A. Wong, "COVID-Net: A Tailored Deep Convolutional Neural Network Design for Detection of COVID-19 Cases from Chest X-Ray Images," Mar. 2020.
[23] X. Xu *et al.*, "Deep Learning System to Screen Coronavirus Disease 2019 Pneumonia," *Appl. Intell.*, vol. 2019, pp. 1–5, Feb. 2020.
[24] C. Zheng *et al.*, "Deep Learning-based Detection for COVID-19 from Chest CT using Weak Label," *medRxiv*, p. 2020.03.12.20027185, Mar. 2020.
[25] S. Chaganti *et al.*, "Quantification of Tomographic Patterns associated with COVID-19 from Chest CT," *ArXiv*, no. 3, Apr. 2020.
[26] F. Shan *et al.*, "Lung Infection Quantification of COVID-19 in CT Images with Deep Learning," Mar. 2020.
[27] D.-P. Fan *et al.*, "Inf-Net: Automatic COVID-19 Lung Infection Segmentation from CT Images," *IEEE Trans. Med. Imaging*, vol. 39, no. 8, pp. 2626–2637, Apr. 2020.
[28] D. Müller, I. S. Rey, and F. Kramer, "Automated Chest CT Image Segmentation of COVID-19 Lung Infection based on 3D U-Net," Jun. 2020.
[29] A. Voulodimos, E. Protopapadakis, I. Katsamenis, A. Doulamis, and N. Doulamis, "Deep learning models for COVID-19 infected area segmentation in CT images," *medRxiv*, p. 2020.05.08.20094664, May 2020.
[30] Q. Yan *et al.*, "COVID-19 Chest CT Image Segmentation -- A Deep Convolutional Neural Network Solution," Apr. 2020.
[31] "'COVID-19 CT segmentation dataset,'" *https://medicalsegmentation. com/covid19/, accessed: 2020-04-11*.
[32] "https://www.sirm.org/category/senza-categoria/covid-19," *https://www.sirm.org/category/senza-categoria/covid-19*.
[33] A. M. Mendrik, E. J. Vonken, A. Rutten, M. A. Viergever, and B. Van Ginneken, "Noise reduction in computed tomography scans using 3-D anisotropic hybrid diffusion with continuous switch," *IEEE Trans. Med. Imaging*, vol. 28, no. 10, pp. 1585–1594, 2009.
[34] O. Ronneberger, P. Fischer, and T. Brox, "U-net: Convolutional networks for biomedical image segmentation," in *Lecture Notes in Computer Science (including subseries Lecture Notes in Artificial Intelligence and Lecture Notes in Bioinformatics)*, 2015, vol. 9351, pp. 234–241.
[35] G. Huang, Z. Liu, L. van der Maaten, and K. Q. Weinberger, "Densely Connected Convolutional Networks," *Proc. - 30th IEEE Conf. Comput. Vis. Pattern Recognition, CVPR 2017*, vol. 2017-January, pp. 2261–2269, Aug. 2016.
[36] K. He, X. Zhang, S. Ren, and J. Sun, "Deep residual learning for image recognition," in *Proceedings of the IEEE Computer Society Conference on Computer Vision and Pattern Recognition*, 2016, vol. 2016-December, pp. 770–778.
[37] Y. Qiu, Y. Liu, and J. Xu, "MiniSeg: An Extremely Minimum Network for Efficient COVID-19 Segmentation," Apr. 2020.